# Sidelobe Level Reduction in the ACF of NLFM Signals Using the Smoothing Spline Method

Roohollah Ghavamirad, Ramezan Ali Sadeghzadeh, and Mohammad Ali Sebt

*Abstract—* **The high level of sidelobes in the autocorrelation function of the nonlinear frequency modulation signal is a challenge. One of the conventional methods to reduce the sidelobe levels is to use the principle of stationary phase. In this method, the frequency function is calculated using a selection window. The signal frequency function cannot be obtained in closed form and numerical methods must be used to find it. This is usually done using the polynomial curve fitting. In this paper, the frequency function of the signal has been obtained using the smoothing spline method. The simulation results show an improvement of 10 dB to 20 dB in the peak sidelobe level of the autocorrelation function of the nonlinear frequency modulation signal compared to the previous methods.**

*Index Terms—* **Peak sidelobe level, stationary phase, polynomial curve fitting, smoothing spline method.**

## I. Introduction

It is not possible to transmit a unmodulated pulse due to its low energy, and on the other hand, if the pulse length is increased, the range resolution required for identifying close targets is significantly reduced. Therefore, the best solution is pulse modulation [1, 2]. By performing modulation, known as pulse compression, the bandwidth and signal-to-noise ratio (SNR) are increased, and range resolution is improved [3, 4]. Linear frequency modulation (LFM) is the most popular pulse compression method, in which the frequency changes linearly over time. Its important advantages are the continuity of the phase and constant signal amplitude, which prevent the mismatches and the loss of SNR at the receiver. However, the presence of strong sidelobes in the autocorrelation function (ACF) is annoying, making it necessary to find a solution to reduce them [5]. Since the Fourier transform of the autocorrelation function corresponds to the power spectral density (PSD), the sidelobes in the autocorrelation function can be reduced by properly shaping the PSD [6]. There are two methods for shaping the PSD.

The authors are with the K. N. Toosi University of Technology, Tehran, Iran (e-mail: sadeghz@eetd.kntu.ac.ir).
Corresponding author: Ramezan Ali Sadeghzadeh

The first method involves weighting the amplitude of the transmitted signal. However, using the variable amplitude requires a linear power amplifier, which is inherently less efficient compared to a saturated power amplifier. Additionally, any mismatch between the receiver and the transmitter can lead to a loss in the SNR [7]. The second method involves weighting the frequency of the signal, which is called nonlinear frequency modulation (NLFM). In this method, the frequency of the signal changes non-linearly over time while its amplitude remains constant. This constancy simplifies the transmitter's implementation and prevents SNR loss, leading to improved performance at lower transmission power. Moreover, the NLFM method significantly reduces sidelobe levels in the autocorrelation function compared to the LFM method [8, 9], although it slightly increases the main lobe width [10, 11]. The first work in the field of NLFM was carried out by Fowle in 1964, and it was based on the stationary phase principle. In 1967, Cook and Bernfeld did further research in this field. The principle of stationary phase states that the power spectral density at a frequency is relatively large if the frequency change rate over time is relatively small [5]. Generally, two design methods have been developed for the NLFM signal. The first method is based on the principle of stationary phase, while the second method involves using explicit functions to express the frequency of the transmitted signal. The second method is used when the frequency changes over time are somewhat known and predictable [12, 13]. Many methods have been proposed to design the NLFM waveform, which are often based on the stationary phase principle. In these methods, the desired PSD for the transmitted signal is first considered, and then, using the stationary phase principle, the group delay function is calculated. The inverse of the group delay function is equivalent to the signal frequency function, but it cannot be expressed in closed form and must be computed numerically. So far, most methods for calculating the inverse of the group delay function have been proposed in various papers, often using polynomial curve fitting. In this paper, curve fitting is performed using the smoothing spline method.

Sections II and III describe the stationary phase method and its implementation using two different windows, respectively. Sections IV and V discuss the polynomial curve fitting and the smoothing spline method, respectively. Section VI presents the simulation results and a comparison of the proposed method with previous methods. Section VII provides the conclusions.

## II. STATIONARY PHASE METHOD

Assume that the NLFM signal in the time domain is given by the following equation [14]:

$$x(t) = a(t)\exp\{j\varphi(t)\}, \quad -\frac{T}{2} \leq t \leq \frac{T}{2} \quad (1)$$

The $a(t)$ and $\varphi(t)$ are the amplitude and phase of the signal $x(t)$, respectively. Additionally, $T$ denotes the pulse length. Let us assume that the Fourier transform of the signal $x(t)$ is represented by $X(f)$ [15]:

$$X(f) = \int_{-\infty}^{\infty} a(t)\exp\{-j[2\pi ft - \varphi(t)]\}dt \quad (2)$$

In general, this integral cannot be expressed in closed form. However, the principle of stationary phase provides a useful approximation for solving it. According to this principle, when the frequency changes slowly over time, the PSD at that frequency will be relatively large. The real part of the signal spectrum is as follows:

$$Re[X(f)] = \int_{-\infty}^{\infty} a(t)\cos[2\pi ft - \varphi(t)]dt \quad (3)$$

If the angle $2\pi ft - \varphi(t)$ changes rapidly compared to the variation of the function $a(t)$, the value of the integral in the adjacent positive and negative parts of the cosine function almost cancel each other. Applying the principle of stationary phase means to take this issue into account and focusing on the points where the angle of the cosine function changes slowly.

$$d[2\pi ft - \varphi(t)]/dt = 0 \quad (4)$$

The integral in (2) can be expanded as a Taylor series around the desired point $t_m$. The first-order term in the angle is zero, as indicated by (4). Therefore, we keep the zeroth-order term in $a(t)$ and the second-order term in the angle. For simplicity, we assume that there is only one stationary phase point.

$$X(f) = a(t_m)\exp\{-j[2\pi ft_m - \varphi(t_m)]\} \quad (5)$$
$$\times \int_{t_m-\delta}^{t_m+\delta} \exp\{j\varphi''(t_m)(t-t_m)^2/2\}dt$$

where $2\delta$ is the distance over which the quadratic approximation for the phase function, according to (2), is valid. The $\varphi''(t)$ is the second derivative of the phase of the signal $x(t)$. By applying a change of variable in Equation (5), the following equation is obtained:

$$X(f) = 2a(t_m)\sqrt{2\pi/|\varphi''(t_m)|}\exp\{-j[2\pi ft_m - \varphi(t_m)]\} \quad (6)$$
$$\times \int_0^{\delta\sqrt{|\varphi''(t_m)|/2\pi}} \exp\{jsgn[\varphi''(t_m)]\pi x^2\}dx$$

In the special case where the upper limit of the integral in Equation (6) is extended to infinity with negligible error, the obtained Fresnel integral can be calculated as follows [15]:

$$X(f) = a(t_m)\sqrt{2\pi/|\varphi''(t_m)|}\exp\{-j2\pi ft_m + \varphi(t_m) \quad (7)$$
$$+ sgn[\varphi''(t_m)]\pi/4\}$$

According to Equation (7), the amplitude of the power spectral density is expressed as follows.

$$|X(f)| = a(t_m)\sqrt{\frac{2\pi}{|\varphi''(t_m)|}} \quad (8)$$

In [14], it is shown that by selecting a suitable window function $X(f)$, the desired signal $x(t)$ can be reconstructed using the principle of stationary phase.

## III. IMPLEMENTING THE STATIONARY PHASE METHOD USING DIFFERENT WINDOW FUNCTIONS

### A. Gaussian window

The Gaussian window can be expressed as the following equation [16]:

$$w(n) = exp\left(-k\left(\frac{n}{2(M-1)}\right)^2\right), \quad |n| \leq \frac{M-1}{2} \quad (9)$$

In (9), $M$ is the window length, and $k$ is a constant positive used to control the sidelobe levels. If $f = nB/(M-1)$, then Equation (9) can be rewritten as follows.

$$w(f) = \exp\left(-k\left(\frac{f}{2B}\right)^2\right), \quad -B/2 \leq f \leq B/2 \quad (10)$$

The integral of $w(f)$ with respect to $f$ cannot be calculated in closed form; however, during the solution process, we encounter a well-known integral called the error function. By applying the boundary conditions, the group delay function can be written as follows.

$$T_g(f) = \frac{T}{2\,erf\left(\sqrt{k}/4\right)}erf\left(\frac{f\sqrt{k}}{2B}\right), \quad -B/2 \leq f \leq B/2 \quad (11)$$

In fact, obtaining $T_g(f)$ in closed form is not possible, but the related data can be generated. Therefore, the inverse calculation of $T_g(f)$ must be performed using numerical methods.

### B. Taylor window

The Taylor window can be expressed as the following Equation.

$$w(n) = 1 + \sum_{m=1}^{n-1} F_m \cos\left(\frac{2\pi mn}{M-1}\right), \quad |n| \leq \frac{M-1}{2} \quad (12)$$

In (12), $F_m = F(m, n, \eta)$ represents Taylor coefficients of order $m$. Moreover, $n$ is the number of sidelobes with equal levels, and $\eta$ is the ratio of the mainlobe level to the peak sidelobe level in the PSD, usually expressed in dB [17]. Assuming $f = nB/(M-1)$, the group delay function can be obtained by integrating and applying the boundary conditions [14], as shown in the following equation:

$$T_g(f) = T\left(\frac{f}{B} + \frac{1}{2\pi}\sum_{m=1}^{n-1}\frac{F_m}{m}\sin\left(\frac{2\pi m f}{B}\right)\right), -B/2 \leq f \leq B/2 \quad (13)$$

The inverse of the group delay function cannot be calculated in closed form and must be determined using numerical methods.

## IV. POLYNOMIAL CURVE FITTING

Curve fitting is a widely used technique in statistics and mathematics for modeling the relationship between dependent and independent variables. One of the most common methods is polynomial curve fitting, where this relationship is represented using a polynomial equation.

Suppose we have a set of data represented as $(x_i, y_i)$ for $i = 0, \ldots, n$, where $x_i$ are the values of the independent variable and $y_i$ are the corresponding values of the dependent variable. The goal of curve fitting is to find a function $f(x)$ that best models the observed data. In the polynomial method, $f(x)$ is defined as a polynomial of degree m [18]:

$$f(x) = a_0 + a_1 x + a_2 x^2 + \cdots + a_m x^m \quad (14)$$

In (14), the $a_i$ values are the polynomial coefficients that must be chosen to minimize the difference between the predicted values and the actual data values. This difference is typically expressed using the sum of squared errors (SSE) as follows.

$$SSE = \sum_{i=0}^{n}(y_i - f(x_i))^2 \quad (15)$$

To minimize the error in Equation (15), numerical methods such as the least squares (LS) method are usually used.

## V. SMOOTHING SPLINE METHOD

The smoothing spline is an effective method for analyzing noisy data and estimating continuous functions. It is suitable for accurately fitting a model to data while ensuring a smooth curve. The main goal is to find a function that fits the data well, while maintaining smoothness and continuity. Smoothing splines are commonly used in nonparametric regression problems, where the exact form of the function is unknown.

Assuming that the observations and independent variable are denoted by $y_i$ and $x_i$ respectively, then the smoothing spline is described by the following equation [18]:

$$y_i = f(x_i) + \epsilon_i \quad (16)$$

In (16), $\epsilon_i$ is random noise with zero mean. The goal is to estimate the function $f(x)$ in such a way that it fits the observed data and is sufficiently smooth. To find a suitable $f(x)$ for the smoothing splines, the following equation is minimized:

$$J = \lambda \int f''(x)^2 dx + \sum_{i=1}^{n}(f(x_i) - y_i)^2 \quad (17)$$

TABLE I
DESIGN PARAMETERS FOR NLFM SIGNAL

| Parameter | Value | Unit |
|---|---|---|
| pulse length | 2.5 | μs |
| | 10 | |
| bandwidth | 100 | MHz |
| sampling rate | 500 | MHz |

In (17), the parameter $\lambda$ determines the trade-off between the fit of the estimated function to the observations and the smoothness of the function. Its value is non-negative. The $f''(x)$ is the second derivative of the $f(x)$.

The first term of Equation (17) indicates the curvature of the estimated function. As this value increases, the estimated function becomes smoother. The second term represents the sum of squared errors, which measures how well the estimated function fit the observed data. Reducing this term improves the alignment between the function and the observed data.

As $\lambda \to \infty$, the focus shifts more toward smoothing the function. However, this may lead to underfitting. As $\lambda \to 0$, the focus shifts toward fitting the function more closely to the observed data. However, this may lead to overfitting.

## VI. SIMULATIONS AND RESULTS

In this section, NLFM signals are designed based on the stationary phase principle using two methods: polynomial curve fitting and smoothing spline. The performance of these methods is evaluated by analyzing their effectiveness in reducing the sidelobes of the autocorrelation function. The relevant design parameters are summarized in Table I.

To evaluate the results, metrics such as the main lobe width (MLW) and the peak sidelobe level (PSL) are used. In earlier studies, the MLW has been calculated at both the -4 dB level (e.g., references [11, 19, 20]) and the -3 dB level (e.g., references [21, 22]). In this paper, the -4 dB level is selected for MLW calculation. The relationship between MLW and PSL is inverse, meaning that as PSL decreases, MLW increases. However, when MLW increases too much, range accuracy is reduced, making it more difficult to distinguish close targets.

Figures 1 to 4 show the autocorrelation function of the desired NLFM signal with different windows, calculated using both polynomial curve fitting and smoothing spline methods. To improve clarity, the time interval from -1 to 1 μs has been selected.

Tables II and III present the PSL and normalized MLW (NMLW) values for two different windows, respectively. The NMLW is defined as the ratio of the main lobe width in the autocorrelation function of the NLFM signal to that of the LFM signal [14].

The smoothing spline method, compared to polynomial curve fitting [16], reduces the PSL by approximately 10 and 20 dB for pulse lengths of 2.5 and 10 μs, respectively. Additionally, it demonstrates an improvement of approximately 10 dB in the PSL compared to the method in [23] for a pulse length of 10 μs. In this method, the greatest reduction in the sidelobe level is achieved with the Gaussian window. The peak sidelobe levels for pulse lengths of 2.5 and 10 μs are -41.69 and -52.46 dB, respectively. Similarly, with polynomial curve fitting, these values are -31.64 and -32.46 dB, respectively. Additionally, in [23], when using the Gaussian window with a pulse length of 10µs, the peak sidelobe level is reported as -42.3 dB. However, the increase in the main lobe width is negligible compared to the reduction in peak sidelobe level.

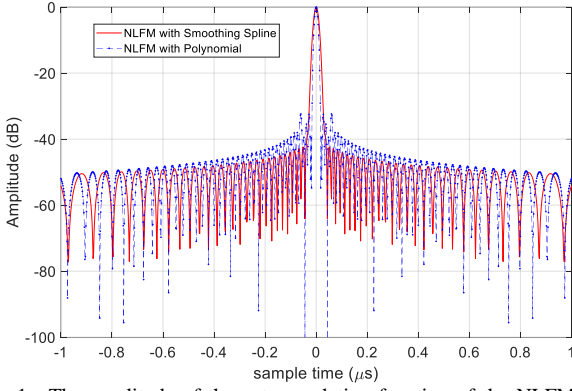

Fig. 1. The amplitude of the autocorrelation function of the NLFM signal using a Gaussian window for a pulse length of 2.5 µs.

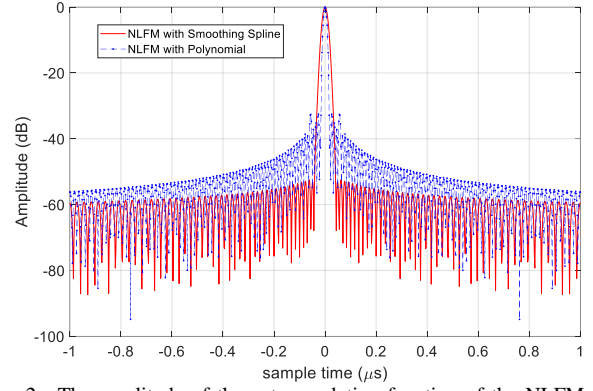

Fig. 2. The amplitude of the autocorrelation function of the NLFM signal using a Gaussian window for a pulse length of 10 µs.

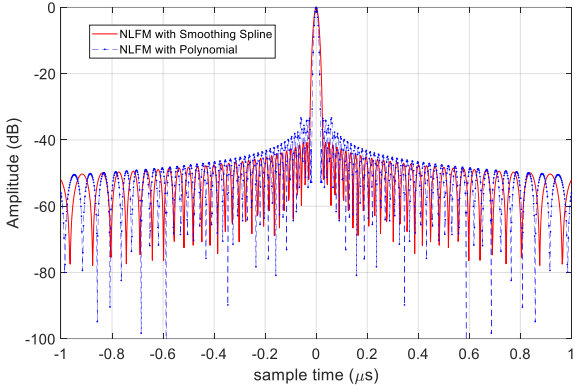

Fig. 3. The amplitude of the autocorrelation function of the NLFM signal using a Taylor window for a pulse length of 2.5 µs.

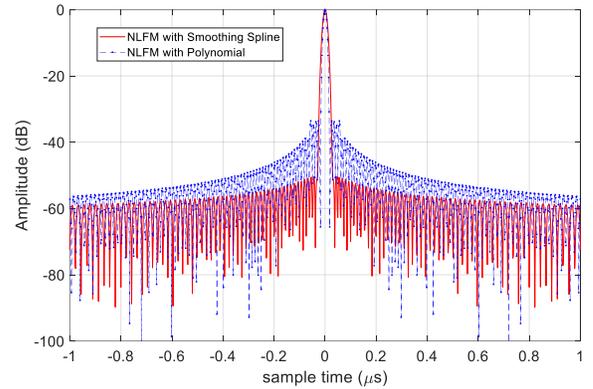

Fig. 4. The amplitude of the autocorrelation function of the NLFM signal using a Taylor window for a pulse length of 10 µs.

TABLE II
COMPARISON OF PSL VALUES FOR TWO DIFFERENT WINDOWS

| Window | Pulse length (µs) | PSL (dB) | | |
|---|---|---|---|---|
| | | Polynomial curve fitting [16] | Ref. [23] | Smoothing spline |
| Gaussian | 2.5 | -31.64 | - | -41.69 |
| | 10 | -32.46 | -42.3 | -52.46 |
| Taylor | 2.5 | -33.34 | - | -40.46 |
| | 10 | -33.57 | -43 | -50.40 |

TABLE III
COMPARISON OF NMLW VALUES FOR TWO DIFFERENT WINDOWS

| Window | Pulse length (µs) | NMLW | | |
|---|---|---|---|---|
| | | Polynomial curve fitting [16] | Ref. [23] | Smoothing spline |
| Gaussian | 2.5 | 1.39 | - | 2.01 |
| | 10 | 1.36 | 1.37 | 2.24 |
| Taylor | 2.5 | 1.36 | - | 1.88 |
| | 10 | 1.35 | 1.31 | 1.83 |

## VII. CONCLUSION

The sidelobe levels in the autocorrelation function of the NLFM signal were significantly reduced using the smoothing spline method compared to polynomial curve fitting. The simulation results showed that for pulse lengths of 2.5 and 10µs, the improvements were approximately 10 and 20 dB, respectively. Although this reduction in sidelobe levels leads to a slight increase in the main lobe width of the autocorrelation function, this increase is negligible compared to the benefits. Furthermore, the sidelobe arrangement in the autocorrelation function using the smoothing spline method is more regular, which enhances range accuracy and improves range resolution. The overall evaluation and conclusion indicate that the efficiency of the smoothing spline method is both significant and impressive. Using this method, the sidelobe level in the autocorrelation function of the NLFM signal can be reduced by up to approximately -53 dB, which is highly applicable in various telecommunication fields, particularly for target discrimination.